\documentclass{article}

\usepackage{PRIMEarxiv}
\usepackage{array}
\usepackage{multirow}
\usepackage{siunitx}
\usepackage[utf8]{inputenc} 
\usepackage[T1]{fontenc}    
\usepackage{hyperref}       
\usepackage{url}            
\usepackage{booktabs}       
\usepackage{amsfonts}       
\usepackage{nicefrac}       
\usepackage{microtype}      
\usepackage{amsmath}
\usepackage{lipsum}
\usepackage{fancyhdr}       
\usepackage{graphicx}       
\usepackage{tabularx}
\usepackage{makecell}
\usepackage{booktabs}
\usepackage{threeparttable}
\usepackage{subfigure}
\usepackage{amssymb}
\graphicspath{{media/}}     
\usepackage{authblk}
\title{Spatial Craving Patterns in Marijuana Users: Insights from fMRI Brain Connectivity Analysis with High-Order Graph Attention Neural Networks}

\author[1]{Jun-En Ding}
\author[1]{Shihao Yang}
\author[2]{Anna Zilverstand}
\author[3]{Kaustubh R. Kulkarni,}
\author[4]{Xiaosi Gu}
\author[1,$\dagger$]{Fegn Liu}

\affil[1]{School of Systems and Enterprises, Stevens Institute of Technology, Hoboken, USA}
\affil[2]{Department of Psychiatry and Behavioral Sciences,
University of Minnesota, Minneapolis, MN 55414, USA}
\affil[3]{Medical Scientist Training Program at the Icahn School of Medicine at Mount Sinai, NY 10027, United States.}
\affil[4]{Psychiatry and Neuroscience and directs the Center for Computational Psychiatry at Mount Sinai in New York City, New York, United States.}
\affil[$\dagger$]{Corresponding author}

\begin{document}
\maketitle

\begin{abstract}
The excessive consumption of marijuana can induce substantial psychological and social consequences. In this investigation, we propose an elucidative framework termed high-order graph attention neural networks (HOGANN) for the classification of Marijuana addiction, coupled with an analysis of localized brain network communities exhibiting abnormal activities among chronic marijuana users. HOGANN integrates dynamic intrinsic functional brain networks, estimated from functional magnetic resonance imaging (fMRI), using graph attention-based long short-term memory (GAT-LSTM) to capture temporal network dynamics. We employ a high-order attention module for information fusion and message passing among neighboring nodes, enhancing the network community analysis. Our model is validated across two distinct data cohorts, yielding substantially higher classification accuracy than benchmark algorithms. Furthermore, we discern the most pertinent subnetworks and cognitive regions affected by persistent marijuana consumption, indicating adverse effects on functional brain networks, particularly within the dorsal attention and frontoparietal networks. Intriguingly, our model demonstrates superior performance in cohorts exhibiting prolonged dependence, implying that prolonged marijuana usage induces more pronounced alterations in brain networks. The model proficiently identifies craving brain maps, thereby delineating critical brain regions for analysis.
\end{abstract}


\section{Introduction}
The issue of substance use disorder (SUD) has become a pressing concern in the United States, with the legalization of non-medical marijuana usage. Marijuana-related disorder accounts for a substantial proportion of individuals seeking treatment for drug use disorders due to the high global prevalence of marijuana use  \cite{connor2021cannabis}.
 Heavy marijuana consumption has been found to impact brain function and cognition ~\cite{hanson2010longitudinal,solowij2010cognitive,crean2011evidence}. In the traditional view, addiction is associated with brain abnormalities in the nucleus accumbens, prefrontal cortex, and amygdala, extensively investigated through neuroimaging techniques~\cite{di1999drug,koob2010neurocircuitry,noel2013neurocognitive,zilverstand2018neuroimaging}. The regional structural or functional changes among patients with SUD may influence addictive behavior by altering reward processing, decision-making, and emotional regulation \cite{abuse2016neurobiology}. However, current research indicates that addiction is not merely a localized brain abnormality but a disruption at the network level. Interacting neural assemblies or brain regions constitute brain networks, and disruptions in connectivity between different regions may crucially impact addictive behavior. Addiction is thus viewed as a network-wide phenomenon with altered functional connectivity patterns, as indicated by network analysis based on fMRI~\cite{volkow2015brain,koob2016neurobiology,everitt2016drug,ma2010addiction}. Recent studies have shown that leveraging fMRI-derived brain network analysis can offer new insights into the relationships between underlying brain connectivity alterations and the manifested characteristics of addictive behavior~\cite{lichenstein2023distinct,betz2022network,ramaekers2022functional}. These characteristics include the age of onset of marijuana use, environmental influences, symptoms of addiction [14], and predictions of marijuana withdrawal and personalized treatment~\cite{lichenstein2023distinct}. 

Researchers have developed several interpretable predictive frameworks based on classical machine learning models to identify individuals at risk of SUD. These frameworks, which utilize algorithms like support vector machines (SVMs) and random forests (RFs), can predict the likelihood of SUD based on different age groups  \cite{jing2020analysis}. The interpretable machine learning frameworks identified risk factors for SUD by investigating gender differences, personality characteristics, and cognitive abilities~\cite{niklason2022explainable,rajapaksha2022bayesian,gadgil2020spatio}.

Constructing brain connectomes using functional or structural neuroimaging modalities has become one of the most pervasive frameworks for brain imaging analysis \cite{cui2022braingb}. In  recent years, graph neural networks (GNNs) have been employed to analyze brain networks for neurological disorder classification and subtyping~\cite{cui2022braingb}. These functional brain networks are constructed from correlation maps of fMRI, where each vertex represents a region of interest (ROI), and edges represent 
the connection strength between different ROIs
~\cite{li2021braingnn,zhu2022joint,liu2023braintgl,kan2023dynamic}. Researchers have utilized a multimodal approach of diffusion tensor imaging (DTI) and fMRI to diagnose mental network disorders  \cite{zhu2022joint}, and have also built functional brain networks to classify autism spectrum disorder (ASD) ~\cite{li2020graph,shao2021classification,zhang2022classification}. Moreover, Schizophrenia can be identified with salient regions based on its topological structure using SVM and Graph Convolutional Network (GCN)~\cite{kipf2016semi}, as demonstrated by six fMRI datasets~\cite{lei2022graph}.

However, fewer studies have conducted functional brain network analysis based on GNN for long-term marijuana users to (1) improve the classification accuracy and (2) delineate the related craving maps or explain alterations of brain networks.

To bridge the research gap, in this study, we propose an interpretable high-order graph attention neural networks (HOGANN) model for analyzing the abnormal brain activities demonstrated in long-term marijuana (LM) users  compared to healthy controls (HC). The main contributions of this study can be summarized as follows: (1) proposing to utilize graph deep learning methods to classify LM users, (2) leveraging the sequential multigraphs in the fMRI time series of marijuana users, (3) integrating the functionality of high-order attention for node relation awareness and message passing into our model, and, (4) we employed analysis across two data sets, successfully identifying significant subnetworks associated with craving maps among LM users, which provided valuable insights into the underlying mechanisms.

\begin{figure*}
\centering
\includegraphics[width=0.9\textwidth]{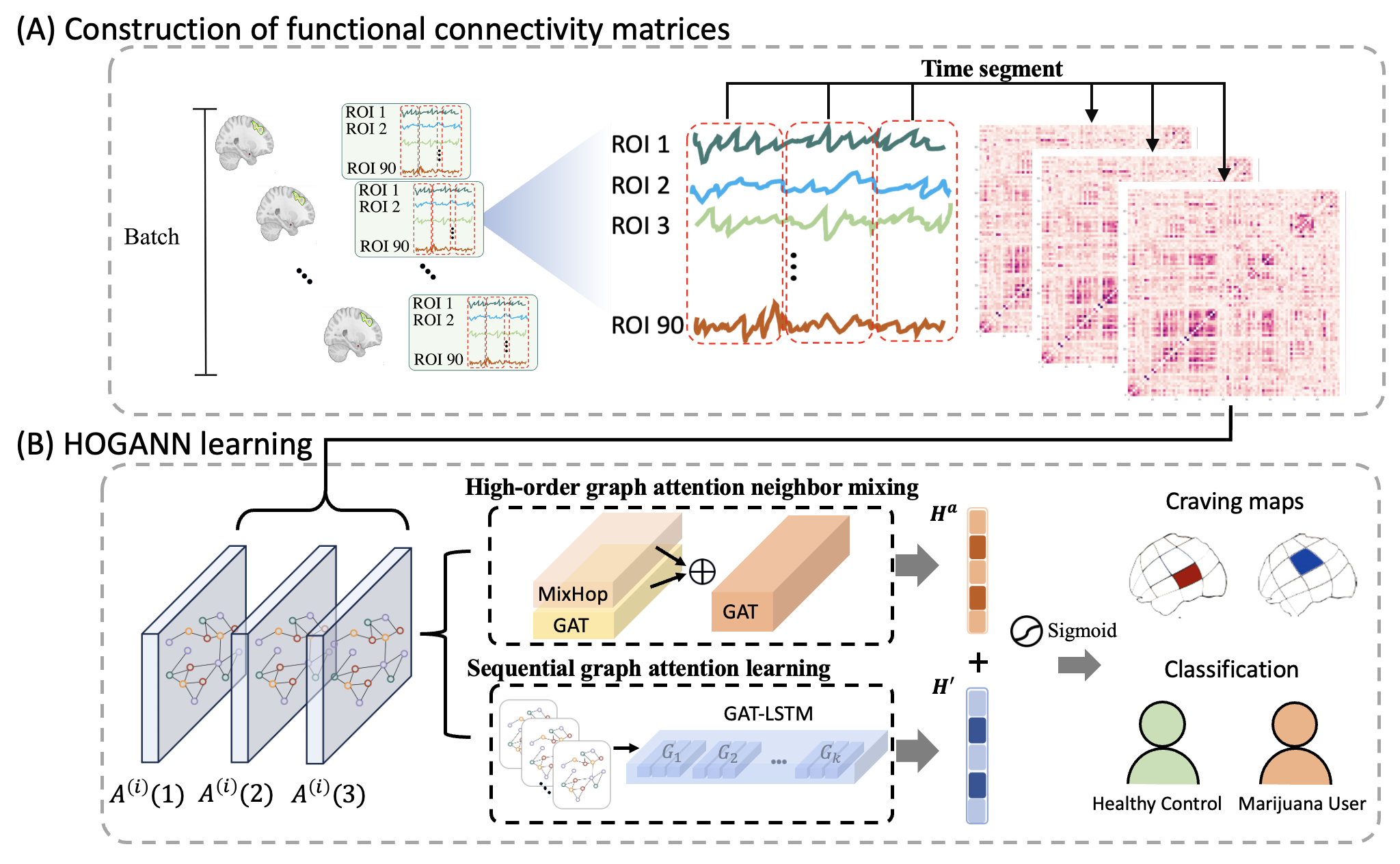}
\caption{The overall framework of the HOGANN utilizes two sub-models to perform fusion learning on fMRI time series. The high-order graph attention neighbor mixing model combines hopping to conduct high-order attention for message passing. Sequential graph attention learning uses GAT-LSTM to learn the temporal sequence graph and discern differences between instances.}
\label{fig:Fig1}
\end{figure*}

\section{Method}
\subsection{Problem Definition}
We define the classification of SUD for each subject as a supervised multigraph classification problem, where each instance graph is constructed from a sliding window of the fMRI time series. 
Specifically, we denote the multivariate time-series data for all individuals as $\{\textbf{X}^{(i)}\}_{i=1}^{N}$ with each individual's data being $\textbf{X}^{(i)} \in \mathbb{R}^{v \times T}$, where $v$ represents the number of ROIs, and $T$ is the time series length. 
We segment these non-overlapping time series into multiple segments, each with a length of $T^{\prime}$. This results in a total of $\left| K \right|$ segments for each subject. Subsequently, we construct the  multigraph $\{\textbf{G}^{(i)}(E,V;t)\}_{t=1}^{K}$ = $\{\textbf{A}^{(i)}(t)\}_{t=1}^{K}$ from the segmented subsequences, where the functional connectivity is based on the k-nearest neighbor (k-NN) graph of each ROI for each subject, where $E$ and $V$ denote edges and vertex (region) set, respectively. Each adjacency matrix $\textbf{A}^{(i)}(t) \in \mathbb{R}^{v \times v}$ corresponds to the $t$-th segment in the $i$-th subject's time series, denoted $\textbf{X}^{(i)}(t), (i \in 1, \cdots, N, t \in 1, \cdots, K)$, and corresponding class labels $Y^{(i)}$ for the multigraph classification task.


\subsection{Model Architecture}
The overall framework of HOGANN, as shown in Fig.~\ref{fig:Fig1} encompasses a multigraph classification architecture with two sub-models: (1) the high-order graph attention neighbor mixing dynamic model, which captures features from multigraphs representations of subject fMRI time series, and (2) the sequential graph attention learning model utilizing graph attention-based long short-term memory (GAT-LSTM) to capture the temporal interdependencies among graphs. The advanced graph attention neighbor mixing module integrates graph neighborhood node message passing to enhance learning of inter-regional relationships of ROIs among individual subjects. Meanwhile, the proposed GAT-LSTM sub-model enhances the extraction of structural features from sequential graphs of interconnected nodes, thereby improving the ability to differentiate between marijuana users and healthy controls. The overall framework enables concatenating features from the latent spaces of the two sub-models leading to improved model performance.

\subsubsection{Two-hop Delta Operator}

We can define injective mapping  Delta Operator $f$ to learn distance between two-hop node features from first-degree neighbors to second-degree neighbors, as follow:

\begin{equation}\label{eq:eq1}
f \left( \sigma \left( \hat{A}X \right) - \sigma \left( \hat{A}^2 X \right) \right),
\end{equation}
where operation can be given any  $\hat{A}$, input features $X$ and activate function $\sigma$. The operator allows the model to represent feature differences among neighbors, enabling it to learn the boundary conditions of the graph.

\subsubsection{High-order graph attention neighbor mixing}
To learn dynamic topological brain networks from multigraphs, we can directly replace the traditional graph convolution layers from the transformation of the two-hop Delta Operation based on the Eq. (\ref{eq:eq1}). We utilize one-hop connections for local graph node message passing with a MixHop layer~\cite{abu2019mixhop}. The formulated equation of the classical $l$-th graph hidden convolutional layer can be defined as follows:

\begin{equation}
\textbf{M}^{(l+1)} = \sigma(\hat{\textbf{A}}\textbf{M}^{(l)}\textbf{W}^{(l)}),
\end{equation}
where normalized adjacency matrix is given by $\hat{\textbf{A}} = \textbf{D}^{-\frac{1}{2}}(\textbf{A} + \textbf{I}_{N})\textbf{D}^{-\frac{1}{2}}$, and $\textbf{A} + \textbf{I}_{N}$ denotes the self-loop connection; $D_{ii}=\sum_{j}A_{ij}$ presents the diagonal degree matrix of $\textbf{A}$, $\textbf{W}^{(l)}$ is a learnable weight matrix in the non-linear activation function $\sigma(\cdot)$. By incorporating the neighborhoods of high-order nodes, the MixHop layer facilitates higher-order latent feature learning enables  learning from their neighbors by incorporating the $j$-th power of the self-adjacency matrix $\hat{\textbf{A}}^{j}$, and the two-hop delta operator learns for various neighborhood distances by subtracting different node features. The MixHop graph convolutional layer can be defined as 
\begin{equation}
\textbf{M}^{(l+1)} = \mathop{\Big\Vert}\limits_{j \in P}\sigma(\hat{\textbf{A}}^{j}\textbf{M}^{(l)}\textbf{W}_{j}^{(l)}),
\end{equation}
where $P$ is a set of integers representing the adjacency matrix's powers and the 
column-wise $\mathbin\Vert$ concatenation of multiple learnable weights. This combination within the graph's convolutional layers indirectly facilitates information propagation to two-hop neighbors, resulting in the matrix 
$\textbf{M}^{(l+1)}= [m_{1},m_{2},...,m_{n}]$.

\begin{figure}
\centering
\includegraphics[width=0.45\textwidth]{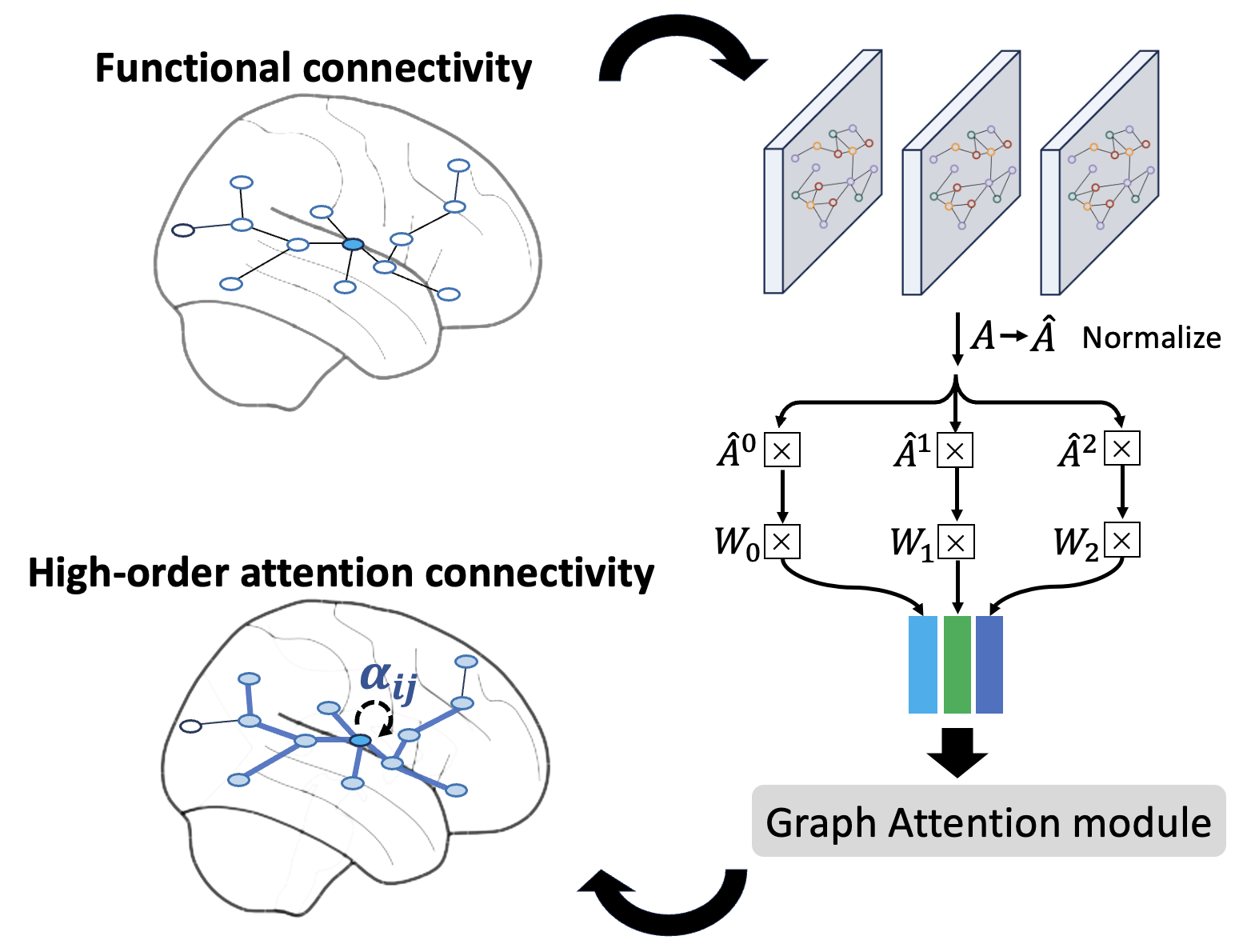}
\caption{The proposed first sub-model contains MixHop and a graph attention layer, wherein the second layer of message passing, concatenating different powers of $\hat{A}^{j}$, can enhance the depth of attention in the neighborhood node connectivity in the pathway.}
\label{fig:Fig2}
\end{figure}

\textbf{Attention with MixHop fusion:}
To effectively integrate temporal graph features and the strength of connectivity paths between nodes through the attention mechanism, we utilized both the MixHop architecture employing high-order message passing and the graph attention network (GAT)~\cite{velivckovic2017graph} module for fusing multigraph node features, as depicted in  Fig. \ref{fig:Fig2}. The GAT layer, with graph attention mechanisms, aggregate neighborhood node features with weighted attention score, while the MixHop layer enhances the connectivity path from source node to high neighbor node of functional connectivity~\cite{yang2018hop}. (Note: first-degree to $k$-degree proximity).
We denote the initial node embedding $ \textbf{H}^{0} = [ \vec{h}^{0}_{1},h^{0}_{2},..., \vec{h}^{0}_{n}]$ from the first graph attention layer and learn the neighbors $\mathcal{N}_{i}$ of ROI nodes $v_{j}$.  We can update the parameterized feature vectors $\vec{h}_{i}$ and the learnable weights $\textbf{W} \in \mathbb{R}^{v \times d}$ using the LeakyReLU function to further calculate the importance attention score with parameterized weight vector $\vec{\textbf{a}}$ across neighboring node $j$ as:

\begin{equation}
\alpha_{ij}=\frac{\mathrm{exp}\left(\mathrm{LeakyReLU}(\vec{\textbf{a}}^T\left[\textbf{W}\vec{h}_{i}  \mathbin\Vert \textbf{W}\vec{h}_{j} \right] \right)}{\sum_{k \in \mathcal{N}_{i} }\mathrm{exp}\left(\mathrm{LeakyReLU}(\vec{\textbf{a}}^T\left[\textbf{W}\vec{h}_{i}  \mathbin\Vert \textbf{W}\vec{h}_{k} \right] \right)},
\end{equation}
The attention mechanism generates an importance attention score $\alpha_{ij}$ for each edge between ROIs, enabling the neighboring  vertex to better focus on the corresponding node at the edge. Subsequently, we can calculate the features representation $\hat{\textbf{H}} =[ \hat{h}_{1},\hat{h}_{2},...,\hat{h}_{v} ]$ of the corresponding $v$ nodes through the attention coefficients after a non-linear transformation, as follows:

\begin{equation}\label{eq:attention}
\hat{h}_{i} =   \sigma \left( \sum_{j \in \mathcal{N}_{i}}  \alpha_{ij}\textbf{W}\vec{h}_{j} \right),
\end{equation}
where $\sigma(\cdot)$ denotes a nonlinear transformation function. In the second layer of the sub-model, we enhance the attention mechanism and the capability for high-order feature fusion within the overall network, as illustrated in Fig. \ref{fig:Fig2}. Furthermore, we concatenate the embeddings $\textbf{M}$ and $ \hat{\textbf{H}}$ from the MixHop and GAT layers, respectively, as the input to the second-layer graph attention layer. However, traditional attention weights for each edge are primarily effective for learning first-order node connectivity, whereas attention MixHop fusion excels at learning second-order neighbor proximity pathways. The concatenated embedding is utilized as the aggregation input for the second-layer $\textbf{GAT}^{(l+1)}$, as follows:
\begin{equation}\label{eq:eq6}
\textbf{H}^{a} =  \textbf{GAT}^{(l+1)}(\left[\textbf{M} \oplus \hat{\textbf{H}} \right]),
\end{equation}
where $\textbf{H}^{a} =\left[ \hat{h}_{1}^{a},\hat{h}_{2}^{a},...,\hat{h}_{k}^{a} \right]$ are output aggregate features vectors, $\oplus$ presents concatenation with corresponding output channel dimensions.

\subsubsection{Attention-driven sequential multigraph learning}

The fMRI signals of the brain's ROIs exhibit variations in dynamic network representation over time $T$ for each instance. The second sub-model was designed to extract dynamic features of functional graphs with cross-sequential dependencies using sequential multigraphs  $\{\textbf{A}^{(i)}(t)\}_{t=1}^{K}$. Traditional LSTM does not consider the graph structure dependencies in longer time sequences. To address this limitation, we propose a modified LSTM architecture that incorporates graph attention mechanisms to capture spatial correlations in node embeddings for multigraph sequences. This attention mechanism effectively enhances the model's ability to learn intrinsic sequential graph structures, particularly in scenarios where extracting interrelated sequential time series at various time points is necessary. Starting from time step $t$, we can formulate the sequence learning from temporal graph representations using the following components: input gate $\textbf{I}_{t}$, forget gate $\textbf{F}_{t}$, output gate $\textbf{O}_{t}$, modulated input $\tilde{\textbf{C}}_{t}$, cell gate $ \textbf{C}_{t}$, and output gate $\textbf{H}_{t}$, as depicted in the following equations:

\begin{align}
    \textbf{I}_{t}&=\sigma(\textbf{W}_{i} \cdot \textbf{GATconv}(\textbf{A}^{(i)}(t))+ \textbf{W}_{i} \cdot \textbf{GATconv}(H_{t-1})) + b_{i} \\[1mm]
    \textbf{F}_{t}&=\sigma(\textbf{W}_{f} \cdot \textbf{GATconv}(\textbf{A}^{(i)}(t)) + 
    \textbf{W}_{f} \cdot \textbf{GATconv}(H_{t-1})) + b_{f} \\[1mm]
    \textbf{O}_{t}&=\sigma(\textbf{W}_{o} \cdot \textbf{GATconv}(\textbf{A}^{(i)}(t)) + 
    \textbf{W}_{o} \cdot \textbf{GATconv}(H_{t-1})) + b_{o} \\[1mm]
    \tilde{\textbf{C}}_{t}&=\tanh(\textbf{W}_{c} \cdot  \textbf{GATconv}(\textbf{A}^{(i)}(t)) + \textbf{W}_{c} \cdot \textbf{GATconv}(H_{t-1})) \\[1mm]
    \textbf{C}_{t} &=\tanh(\textbf{I}_{t} \cdot \tilde{\textbf{C}}_{t} + \textbf{F}_{t} \cdot \textbf{C}_{t-1} ) \\[1mm]
    \textbf{H}_{t} &= \textbf{O}_{t} \cdot \tanh(\textbf{C}_{t})
\end{align}
where $\textbf{W}_{i}$, $\textbf{W}_{f}$, $\textbf{W}{o}$ and $\textbf{W}_{c}$ represent learnable weights at different gates and bias vectors $b_{i}$, $b_{f}$, and $b_{o}$. Notably, we introduce the graph attention convolution layer \textbf{GATconv}$(\cdot)$ to encode sequential graphs $\textbf{A}^{(i)}(t)$ in each LSTM gate. We can define $l$-th \textbf{GATconv}$(\cdot)$ layer as:

\begin{equation}\label{eq:eq5}
\textbf{H}_{t}^{(l+1)}=\textbf{GATconv}(\textbf{A}^{(i)}(t))=\sigma(\textbf{A}^{(i)}(t)\textbf{H}^{(l)}_{t}\textbf{W}^{(l)}_{t}),
\end{equation}
where $\textbf{H}^{(l+1)}_{t}$ is the output features matrix calculated after the $l$-th hidden layer of graph  attention  with trainable weights $\textbf{W}^{(l)}_{t}$.
By taking the output from the GAT-LSTM layer, we obtain the final graph representation for each instance:
\begin{equation}
h^{\prime}_{1},h^{\prime}_{2},...,h^{\prime}_{k}=\textbf{GAT-LSTM}(h_{1},h_{2} ,...,h_{k}), 
\end{equation}
where $h^{\prime}_{1},h^{\prime}_{2},...,h^{\prime}_{k}$ are latent vectors in $t$-th time step from graph attention LSTM modeling as illustrated in Fig. \ref{fig:Fig1}. Then, we can combine temporal graph features to generate instance representation $\textbf{H}^{\prime} = \left[h_{1}^{\prime}, h_{2}^{\prime}, h_{3}^{\prime},..., h_{k}^{\prime} \right]$. To better aggregate two sub-model embeddings, we first combine the representations $\textbf{H}^{a}$ and $\textbf{H}^{\prime}$, as follows:
\begin{equation}\label{eq:eq15}
\textbf{H}^{c}=  \frac{1}{2}(\textbf{H}^{a} + \textbf{H}^{\prime}). 
\end{equation}
where $\textbf{H}^{c}$ represents final fused node embedding, which is a combination of outputs from the two sub-models and is utilized to integrate information contributed by both perspectives provided by these sub-models.

\subsubsection{Brain network reconstruction and weighted functional connectivity}
Functional connectivity assesses brain connections between various regions. To improve brain network connectivity among different ROI/nodes with spatial relationships, we calculate node  $i$ features representation using a mixhop attention mechanism by aggregating fused node features from Eq. \ref{eq:eq15}. Subsequently, we perform graph mean pooling for $i$th node embeddings, which can be expressed as:

\begin{equation}\label{eq:eq16}
L^{c}_{i} = \frac{1}{\left| \mathcal{V}\right|} \sum_{j \in \mathcal{V}} H^{c}_{j},
\end{equation}
where $L^{c}_i \in \mathbb{R}^{1 \times d}$ can be expressed features maps for each ROI, we utilize the final output node embedding, denoted by $\textbf{L}^{c} = \left[L^{c}_{1},L^{c}_{2},...,L^{c}_{v}  \right]$ for brain network reconstruction. This is accomplished through an inner product operation, as depicted in Eq. (\ref{eq:eq14}):
\begin{equation}\label{eq:eq14}
\textbf{B}_{R} = \textbf{L}^{c} \cdot (\textbf{L}^{c})^{{T}},
\end{equation}
where $\textbf{B}_{R} \in \mathbb{R}^{v \times v}$ represents the global brain reconstruction matrix for ROIs.  To enhance the structural consistency of ROIs across subjects in functional connectivity analysis, we employ element-wise multiplication with the reconstruction ROI matrix $\textbf{B}_{R}$. This operation is applied to each subject's connectivity matrix $\textbf{A}^{(i)}(t)$ at time step $t$, where $i$ represents the subject index and $t$ denotes the temporal segment. This computation is shown in Eq. (\ref{eq:fc_w}), and the subject-specific weighted functional connectivity matrices can be expressed as:
\begin{equation}\label{eq:fc_w}
\mathbf{W}_{fc}^{(i)}(t) = \mathbf{A}^{(i)}(t) \odot  \sigma(\textbf{B}_{R}),
\end{equation}
where $\sigma(\cdot)$ is the sigmoid function, which normalizes $\textbf{B}_{R}$ to a range between 0 and 1, and $\odot$ denotes the element-wise product. Subsequently, we average these weighted functional connectivity matrices over time as follows:

\begin{equation}\label{eq:avg_fc}
\mathbf{\Bar{W}}_{fc}^{(i)} = \frac{1}{N} \sum_{t=1}^{N} \mathbf{W}_{fc}^{(i)}(t),
\end{equation}
where $\mathbf{\Bar{W}}_{fc}^{(i)}$ represents the final average weighted functional connectivity (AWFC) matrix. The advantage of constructing an AWFC matrix is its ability to capture an overall weighted representation of the brain network. This representation enables further analysis of degree centrality (DC) and network community differences between distinct groups based on edge density analysis.

\subsection{Objective function}

We can employ the output embeddings from Eq. (\ref{eq:eq6}) and Eq. (\ref{eq:eq15}) throughout the training phase. We then use Eq. (\ref{eq:eq20}) and Eq. (\ref{eq:eq21}) to execute a binary classification task, incorporating the final non-linear transformation $\sigma(\cdot)$ as follows:
\begin{equation}\label{eq:eq20}
\hat{\textbf{Y}}_{c}=\sigma(\textbf{W}_{c}\textbf{H}^{c} + b_{c}),
\end{equation}
\begin{equation}\label{eq:eq21}
\hat{\textbf{Y}}_{MixGAT}=\sigma(\textbf{W}_{a}\textbf{H}^{a} + b_{a}),
\end{equation}
where $\textbf{W}_{c}$ and $\textbf{W}_{a}$ are the trained parameter weights from the two sub-models, while $b_{c}$ and $b_{a}$ represent their respective biases. Subsequently, our objective is to minimize the overall loss function derived from the losses of two sub-models. The first sub-model pertains to the MixHop attention loss, denoted as $\mathcal{L}_{MixGAT}$, while the second sub-model involves the combination fusion model loss denoted as $\mathcal{L}_{c}$. The combined loss can be expressed as:

\begin{equation}\label{eq:eq19}
\mathcal{L} = \mathcal{L}_{MixGAT}(\textbf{Y},\hat{\textbf{Y}}_{MixGAT}) + \mathcal{L}_{c}(\textbf{Y},\hat{\textbf{Y}}_{c}).
\end{equation}
where $\textbf{Y}$ represents the ground truth label, while $\hat{\textbf{Y}}_{MixGAT}$ and $\hat{\textbf{Y}}_{c}$ represent the class predictions for marijuana users and healthy controls, respectively.  By combining $\mathcal{L}_{MixGAT}$ and $\mathcal{L}_{c}$, we can better leverage the fusion information in the total loss $\mathcal{L}$ not just sub-model loss. We will discuss the impact of loss on HOGANN overall prediction performance in more detail in Section \ref{sec:Objective_function}.

\begin{table}
\centering
\caption{Summary statistics of demographics in Marijuana-323 and HCP}
\begin{tabular}{>{\centering\arraybackslash}m{1.2cm}>{\centering\arraybackslash}m{1.9cm}>{\centering\arraybackslash}m{1.9cm}>{\centering\arraybackslash}m{1.9cm}}
\toprule
Cohort & Dataset & \makecell{Age \\ (Mean ± SD)} & \makecell{Gender \\ (Female/Male)} \\
\midrule
\multirow{2}{*}{HC} & Marijuana-323 (n=128) & 30.06 ± 10.47 & 73/55 \\
& HCP (n=991) & 28.82 ± 3.72 & 567/424 \\
\midrule
\multirow{2}{*}{LM} & Marijuana-323 (n=195) & 27.23 ± 7.73 & 73/122 \\
& HCP (n=100) & 28.45 ± 3.49 & 26/74 \\
\hline
\hline
\multirow{2}{*}{Significance} & Marijuana-323 & $t=2.79^{\star}$ & $\chi^2=11.20^{\star\star}$ \\
& HCP & $t=0.95$ &  $\chi^2=34.42^{\star\star}$\\
\bottomrule
\end{tabular}
\label{table:demographics}
\begin{tablenotes}
        \footnotesize
        \raggedright
        \item  Note that: “$\star\star$” and “$\star$” denote statistical significance in demographic characteristics (i.e., Age and Gender) at the $0.001$ and $0.05$ levels respectively. The $t$ represents the paired $t$-test, and $\chi^2$ represents the chi-square test.
\end{tablenotes}
\end{table}

\subsection{Datasets}

\begin{itemize}
      \item \textbf{Marijuana-323:} In this study, we extended the prior work of Kaustubh R. et al~\cite{kulkarni2023interpretable,filbey2016fmri,filbey2009marijuana}. The Institutional Review Board (IRB) of this study was exempted by the Stevens Institute of Technology Review Board review committee. The research involved the collection of data from two datasets of fMRI images comprising 125 and 198 participants, recruited either from the non-medical marijuana-seeking community or from hospitalized patients. Participants were required to withdraw from marijuana for 72 hours and then asked to complete a cue-elicited craving task during an MRI scan. After the MRI scans, participants completed the marijuana craving questionnaire~\cite{heishman2001marijuana}, marijuana withdrawal checklist, and marijuana problem investigation~\cite{budney1999marijuana}. The data underwent reprocessing from 3T imaging fMRI data, and an average time series of 90 ROIs was calculated for each subject based on the Stanford atlas ~\cite{shirer2012decoding}. In our final training and  testing, there were 195 long-term marijuana users and 128 healthy control individuals.

      \item \textbf{HCP dataset:} To assess the effectiveness of HOGANN, we evaluated an external dataset sourced from the human connectome project (HCP) S1200 \cite{van2013wu}, comprising 1096 rs-fMRI scans of young adults. The S1200 constitutes the major releases from 2012 to 2015 from Washington University in St. Louis, Missouri, United States. The HCP dataset
     encompasses 598 individuals who have used marijuana,
     with an average age of 28.76 ± 3.69, and 493 individuals who have never used marijuana, with an average age of 28.82 ± 3.72. This data preparation utilized each subject's first session and excluded 5 rs-fMRIs containing fewer than 1200 frames. The cortical surfaces of each subject were partitioned into 22 brain regions serving as ROIs ~\cite{glasser2016multi}, and blood oxygenation level-dependent (BOLD) time series signals were normalized using z-scores ~\cite{gadgil2020spatio}. 
\end{itemize}

\section{Results}

Table \ref{table:demographics} displays summary statistics of the demographics for two datasets. In our experimental setup, the Marijuana-323 and HCP datasets were divided into training and testing sets in an 80/20 ratio. The Marijuana-323 dataset comprised 258 training subjects and 65 testing subjects, while the HCP dataset comprised 876 training subjects and 218 testing subjects.

\begin{table}[t]
\caption{Evaluating classification performance with five-fold cross-validation. We computed the most competitive baseline for each method. We compared the second-best methods denoted by "$\blacktriangle$" and calculated the improvement rate, denoted as "Improv. (\%)". Here, "$\star$" denotes statistical significance at the 0.05 level using a two-sample t-test.}
\centering
\resizebox{\textwidth}{!}{
\begin{tabular}{|l|c|c|c|c|c|c|c|c|}
\hline
 & \multicolumn{4}{c|}{\textbf{Marijuana-323 (90 ROIs)}} & \multicolumn{4}{c|}{\textbf{HCP (22 ROIs)}} \\
 \hline
\textbf{Model} & \textbf{AUC} & \textbf{Acc. (\%)} & \textbf{Prec. (\%)} & \textbf{Rec. (\%)} & \textbf{AUC} & \textbf{Acc. (\%)} & \textbf{Prec. (\%)} & \textbf{Rec. (\%)} \\
\hline
LR + L1 & 0.71$\pm$0.08 & 73.4$\pm$7.5 & 74.3$\pm$7.5 &  73.5$\pm$7.5 & 0.62$\pm$0.02 & 62.0$\pm$1.9 & 62.0$\pm$5.6 & 68.5$\pm$17.1 \\
LR + L2 & 0.72$\pm$0.09 & 74.4$\pm$8.2 & 75.5$\pm$9.0 & 74.4$\pm$8.3 & 0.57$\pm$0.04 & 57.1$\pm$4.1 & 66.9$\pm$14.3 & 53.9$\pm$35.9 \\
SVM + L1 & 0.68$\pm$0.08 & 69.7$\pm$7.5 & 73.6$\pm$9.5 &  73.1$\pm$9.0 & 0.64$\pm$0.01 & 63.2$\pm$0.8 & 64.1$\pm$1.2 & 64.1$\pm$1.2 \\
SVM + L2 & 0.71$\pm$0.10& 73.0$\pm$8.9 & 73.6$\pm$9.5 & 73.1$\pm$9.0 & 0.64$\pm$0.01 & 63.7$\pm$0.9 & 63.6$\pm$1.0 & 62.6$\pm$1.9 \\
XGboost &  0.65$\pm$0.01 & 67.6$\pm$8.3 & 66.6$\pm$14.4 & 67.6$\pm$8.3 & 0.65$\pm$0.11 & 67.6$\pm$8.3 & 66.6$\pm$14.4 &  67.6$\pm$8.3 \\

\hline
GCN & 0.77$\pm$0.01 & 68.4$\pm$4.5 & 75.7$\pm$8.4 & 74.8$\pm$11.8 & 0.66$\pm$0.06 & 68.0$\pm$2.6 & 69.7$\pm$2.4 & 74.0$\pm$3.2 \\
GAT & 0.79$\pm$0.04  & 72.1$\pm$3.8 & 75.3$\pm$6.2 & 81.7$\pm$9.4 & 0.67$\pm$0.02 & 72.7$\pm$1.6 & 73.1$\pm$0.9 & 75.1$\pm$4.9 \\
GraphSAGE & 0.77$\pm$0.00 & 64.6$\pm$5.3 & 73.6$\pm$5.6 & 68.2$\pm$21.2 & 0.74$\pm$0.01 & 67.5$\pm$2.3 & 68.4$\pm$1.4 & 67.4$\pm$8.5 \\
BrainGB & 0.72$\pm$0.04  & 66.0$\pm$2.2 & 65.5$\pm$2.7 & 64.1$\pm$2.8 & 0.62$\pm$0.01 & 59.1$\pm$0.7 & 59.0$\pm$1.2 & 57.4$\pm$0.9 \\

BrainUSL & 0.64$\pm$0.19 & 62.7$\pm$9.6& 71.3$\pm$9.3 & 73.3$\pm$30.0 & 0.48$\pm$0.26  & 73.2$\pm$12.0 & 73.0$\pm$9.4 & 79.5$\pm$16.0 \\

Graph-Transformer & 0.75$\pm$0.04 & 70.9$\pm$3.5 & 75.1$\pm$5.4 & 77.9$\pm$2.4 & 0.76$\pm$0.06 & 72.4$\pm$2.6 & 72.4$\pm$4.8 & 77.8$\pm$5.2 \\

BrainGSL & 0.80$\pm$0.02$^{\blacktriangle}$ & 74.3$\pm$2.8$^{\blacktriangle}$  & 77.5$\pm$2.2$^{\blacktriangle}$  & 80.7$\pm$6.5$^{\blacktriangle}$  & 0.77$\pm$0.06$^{\blacktriangle}$  & 74.5$\pm$9.6$^{\blacktriangle}$  & 74.3$\pm$6.0$^{\blacktriangle}$  & 81.3$\pm$18.3$^{\blacktriangle}$  \\
\hline
HOGANN (w/o GAT-LSTM) & 0.80$\pm$0.02 & 74.7$\pm$2.8 & 77.3$\pm$3.1 & 82.5$\pm$7.0 & 0.74$\pm$0.06 & 73.2$\pm$5.7 & 74.5$\pm$5.9 & 78.6$\pm$8.0 \\
HOGANN (w/o MixHop) & 0.78$\pm$0.03 & 70.1$\pm$4.4 & 77.3$\pm$3.0 & 83.8$\pm$4.5 & 0.79$\pm$0.04 & 72.7$\pm$5.3 & 73.6$\pm$5.0 & 79.0$\pm$5.3 \\
HOGANN & \textbf{0.86$\pm$0.03} & \textbf{79.2$\pm$3.6}& \textbf{79.7$\pm$4.0} & \textbf{87.9$\pm$1.3}& \textbf{0.85$\pm$0.04} & \textbf{78.5$\pm$2.2} & \textbf{81.2$\pm$3.4} & \textbf{83.0$\pm$3.4} \\
\hline

Improv. (\%) & 7.5\% & 6.5\% & 2.8\% & 8.9\% & 10.3\% & 5.3\% & 9.5\% & 2.0\% \\
\hline
Significance & $\star$  &    &  &  $\star$ &  $\star$ & $\star$ & $\star$ &  \\
\hline
\end{tabular}
}
\label{table:overall_table}
\end{table}

\subsubsection{Classical methods and graph-level classification}

The experiment results of the proposed HOGANN, as shown in Table \ref{table:overall_table}, 
illustrate the model performance across different approaches in the marijuana-323 (90 ROIs) and HCP (22 ROIs) datasets. We first used ML methods as the baseline, our experiments reveal that among the ML methods: LR + L1, LR + L2, SVM + L1, SVM + L2 and XGboost. Particularly, LR + L2 achieved the highest classification accuracy at 74.4\% in Marijuana-323, as depicted in Table \ref{table:overall_table}. However, the same ML methods did not yield superior results in the HCP dataset. These classification results across the two datasets indicate the presence of nonlinear structures among brain regions in the fMRI scans of LM users, which ML methods fail to discern based on functional connectivity features. Finally, to evaluate the graph topology structure of fMRI data, we reasonably compare graph-based deep learning with our proposed HOGANN, such as GCN, GAT, GraphSAGE, and BrainGB as baseline models. Table \ref{table:overall_table} shows that GAT achieve better area under the curve (AUC) scores than ML methods in the marijuana-323 dataset. The experimental results from GAT indicate that utilizing only the one-hop neighborhood of local nodes achieves an AUC of 0.79 in the Marijuana-323 dataset but fails to perform adequately in the HCP dataset (AUC=0.67). Meanwhile, BrainGB and GraphSAGE with mini-batch node feature sampling methods also fail to outperform GAT, which has an attention mechanism, in terms of performance.

\subsubsection{Compare with SOTA models}

Furthermore, we conduct a fair comparison between our HOGANN and more complex graph deep learning architectures, including Graph-Transformer, BrainUSL, and BrainGSL) on two datasets, using a fixed learning rate of 0.01. Based on the experimental results, BrainUSL, an unsupervised framework, exhibits inferior prediction performance on the Marijuana-323 and HCP datasets compared to the ML method when evaluated using AUC scores. We can reasonably explain that unsupervised models are prone to generalization failures when applied across different datasets, leading to poor model performance. Furthermore, we compared the second-best model, BrainGSL, with self-supervised learning to our HOGANN. The comparison resulted in significant improvements: an 7.5\% increase in AUC on the Marijuana-323 dataset and a 10.3\% increase on the HCP dataset. Additionally, on the HCP dataset, we observed improvements of 5.3\% in accuracy and 9.5\% in precision. However, these models typically concentrate solely on the topological information extracted from the entire fMRI time series and do not consider the dynamic topological structure in different time intervals. To address these limitations and validate our approach, we present comprehensive ablation studies in the following section \ref{sec:Objective_function}.

\begin{figure}
\centering
\includegraphics[width=0.8\textwidth]{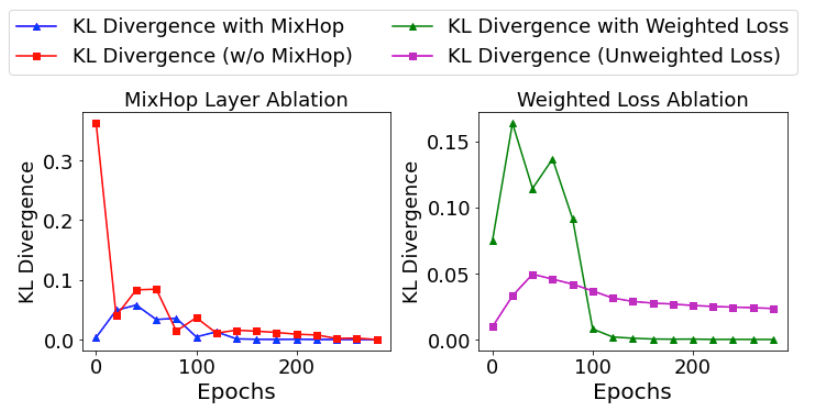}
\caption{The impact of KL divergence ablation studies on mixed embeddings $\textbf{H}^{a}$ and $\textbf{H}^{\prime}$: (Left) Comparing KL divergence with and w/o MixHop. (Right) Comparing KL divergence with weighted loss versus unweighted loss.}
\label{fig:KL}
\end{figure}

 \begin{figure*}
\centering
\includegraphics[width=0.95\textwidth]{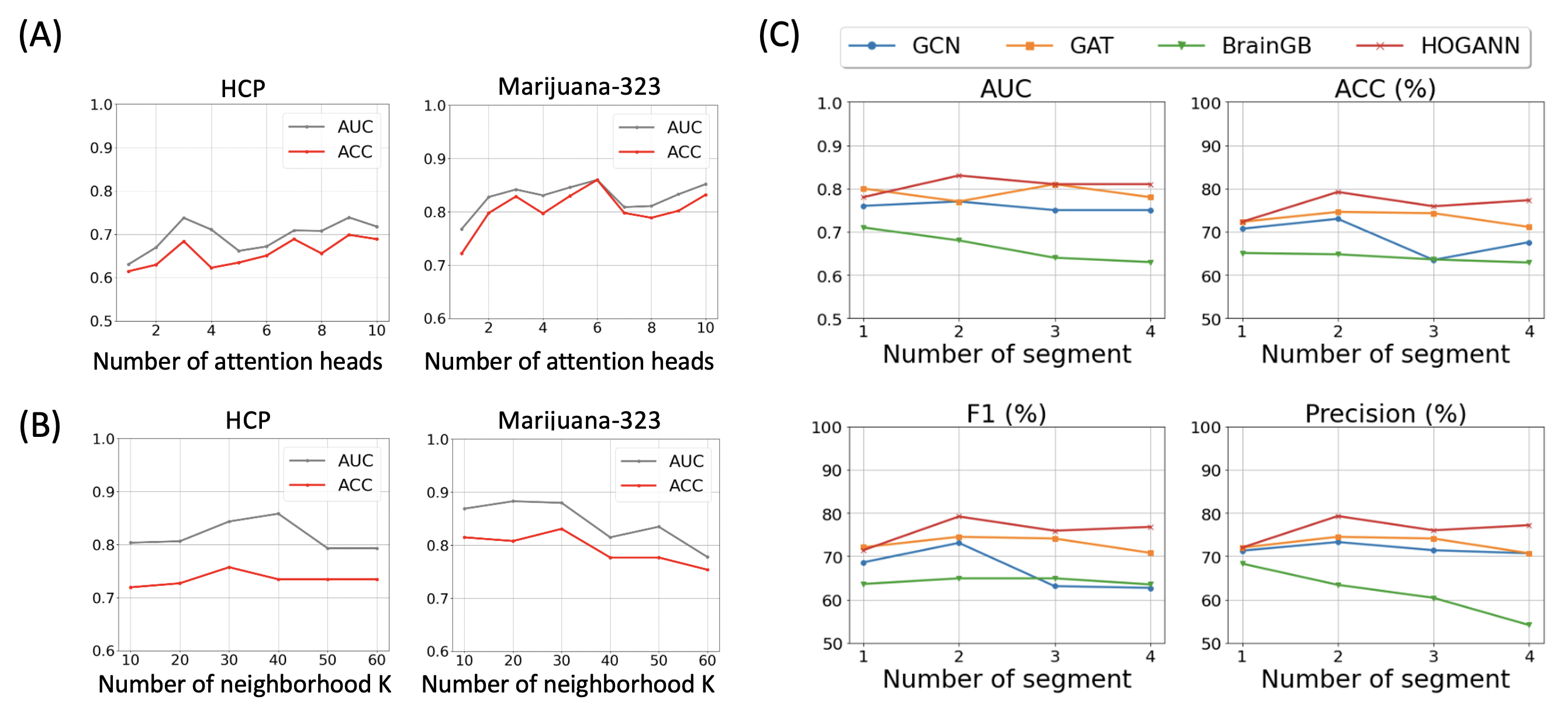}
\caption{The ablation analysis of HOGANN classification in LM and HC classification. Panel (A) demonstrates the performance changes with multiple attention heads, while Panel (B) illustrates the classification performance changes when the neighborhood K in of the HOGANN is varied. Panel (C) comparing different models based on their classification performance on the marijuana-323 with varying multi-segment time window sizes.} 
\label{fig:attention head}
\end{figure*}

\begin{figure}
\centering
\includegraphics[width=0.8\textwidth]{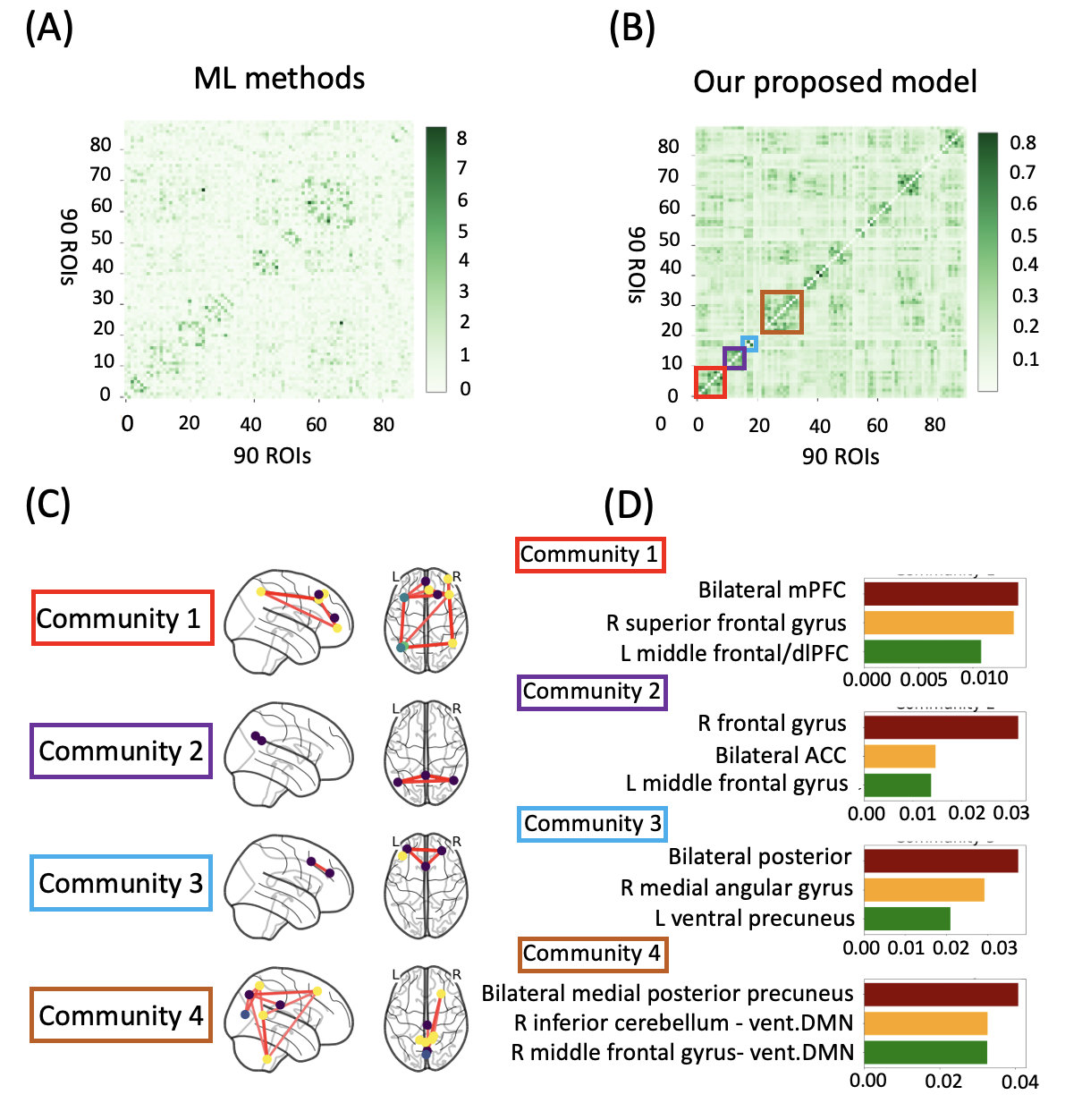}
\caption{A comparison of machine learning methods with our HOGANN in identifying the optimal community clustering. In Panel (A) and (B), we compare community detection in weighted functional connectivity matrices using machine learning and our HOGANN, where we map the most significant four communities of the LM group to their corresponding brain regions of the connectivity network. Based on Panel (C) for the LM group, we depict four communities corresponding to the colored boxes, illustrating the connectivity of the most active brain regions in Panel (D).}
\label{fig:community_cluster}
\end{figure}

\begin{table}
\centering
\caption{The ablation experiment in loss comparison}
\label{your-table}
\setlength{\tabcolsep}{3pt} 
\begin{tabular}{@{}lcccccccc@{}}
\toprule
& \multicolumn{4}{c}{Marijuana-323 (90 ROIs)} & \multicolumn{4}{c}{HCP (22 ROIs)} \\
\cmidrule(r){2-5} \cmidrule(l){6-9}
Strategies & AUC  & Accuracy (\%) & Precision (\%) & Recall (\%) & AUC & ACC (\%) & Precision (\%) & Recall (\%) \\ 
\midrule
$\mathcal{L}_{c}$ & 0.79 & 76.9 & 76.7 & 76.9 & 0.66 & 63.9 & 63.9 & 63.9 \\
$\mathcal{L}_{MixGAT}$ & 0.79 & 73.9 & 73.7 & 73.9 & 0.62 & 68.0 & 68.3 & 68.0 \\
$\alpha \cdot \mathcal{L}_{MixGAT} + (1-\alpha) \cdot \mathcal{L}_{c}$ & 0.79 & 75.4 & 75.4 & 75.4 & 0.59 & 62.5 & 62.5 & 62.5 \\
$\mathcal{L}_{c} +\mathcal{L}_{MixGAT}$ & \textbf{0.84} & \textbf{78.5} & \textbf{78.3} & \textbf{78.5} & \textbf{0.70} & \textbf{69.1} & \textbf{69.4} & \textbf{69.1} \\
\bottomrule
\end{tabular}
\label{table:loss_table}
\end{table}

\subsection{Ablation studies}\label{sec:Objective_function}

In our ablation experiments, we observed that utilizing MixHop or GAT-LSTM independently yielded superior classification performance improvements across various metrics in the LM and HC groups compared to most graph-based model baselines (such as GCN, GAT, GraphSAGE, and BrainGB). However, we noted that the simultaneous integration of higher-order attention mixing and dynamic sequential graph approach enables the preservation of high-order attention mechanisms in overall brain network learning for more effective messaging. Our research conducts the following ablation experiments on HOGANN's hyperparameters, loss function, and the number of segments in the fMRI time series:

\begin{enumerate}

    \item  We evaluated model performance solely using loss function $\mathcal{L}_{c}$. Our experiments showed that HOGANN, which relied on a single model, achieved AUCs of 0.79 and 0.66 on the two datasets shown in Table \ref{table:loss_table}. Notably, using only the $\mathcal{L}_{MixGAT}$ loss for high-order attention yielded similar performance to $\mathcal{L}{c}$ alone but with lower accuracy, precision, and recall in the marijuana-323 dataset. This suggests that the $\mathcal{L}_{c}$ loss effectively leverages the fused embedding effect of $\textbf{H}^{c}$. Although combining loss functions might seem beneficial, our experiment with a weighted sum loss ($\alpha=0.3$) of $\mathcal{L}{c}$ and $\mathcal{L}_{MixGAT}$ did not improve performance. Instead, HOGANN adds the two losses significantly improved results, achieving an AUC of 0.84, accuracy of 78.5\%, precision of 78.3\%, and recall of 78.5\% in marijuana-323. Similar improvements in performance were seen in HCP.

    \item In Fig. \ref{fig:KL}, we illustrate the Kullback-Leibler divergence (KLD) of two embeddings $\textbf{H}^{a}$ and $\textbf{H}^{\prime}$ from the HOGANN sub-models during the training process. Fig. \ref{fig:KL} (left) depicts that HOGANN, when combined with MixHop, effectively stabilizes in KLD convergence of the two embeddings' distributions during training process. Additionally, we also analyze Fig. \ref{fig:KL} (right), which shows that the weighted losses exhibit more significant variation in the early stages of training. From our experiments, we observe that the weighted loss disperses the joint embedding of the two sub-models. The weighting also disperses the convergence of the individual sub-models.

    \item In Fig. \ref{fig:attention head}, we increase the number of graph attention layers with multi-head to capture long-range dependencies and local neighborhood graph structure, as illustrated in Fig. \ref{fig:attention head} (A), and (B). From the results, it can be observed that HOGANN in marijuana-323 and HCP substantially improved classification performance. This indicates that increasing the number of attention heads can enable a broader learning of node embeddings for the high-order connectivity of each vertex.
    \item To evaluate classification  based on varying lengths of fMRI time series, our experiments illustrate the impact of segmenting full-time series and subsequences into multigraphs with different window sizes $T^{\prime}$ in marijuana-323, as illustrated in Fig. \ref{fig:attention head} (C). Our training strategy employs four-time windows of length $T^{\prime} \in$ \{0, 100, 300, 600\} for training and testing. As depicted in Fig. \ref{fig:attention head} (C), it is evident that the AUC, accuracy, F1, and precision exhibit substantial reductions as the number of segmented fMRI signals increases (segment $>$ 3) in the marijuana-323. However, GAT demonstrates a more stable performance when segmenting two or three times compared to GCN. HOGANN showcases increasingly robust classification performance in our experiments as the number of segments increases. Specifically, under the optimal condition of segmenting fMRI signals twice,  HOGANN achieves an AUC of 0.83, accuracy of 79.2\%, F1 of 79.26\%, and precision of 79.3\%.
    
    \item We analyzed the HCP dataset characterized by marijuana dependence (Mj) and those who have previously used marijuana, conducting predictions across different age groups.  As shown in Table \ref{tab:Age_group}, the experimental results indicate that for the group with marijuana dependence, our HOGANN predicts the results more accurately in all age groups compared to the group that has only used marijuana. This indicates stronger deviations in neural networks in individuals with marijuana dependence versus marijuana users as compared to healthy control participants.

\end{enumerate}

\begin{table}
\centering
\caption{Classification Performance by Age Group in HCP Dataset}
\label{tab:Age_group}
\begin{tabular}{@{}llcc@{}}
\toprule
Condition      & Age Group & AUC  & Accuracy (\%) \\ \midrule
\multirow{3}{*}{Marijuana Dependence } & 22-25    & 0.89 & 92.4       \\
                               & 25-30      & 0.90 & 86.4       \\
                               & 30-37      & 0.99 & 87.9       \\ \midrule
\multirow{3}{*}{Ever Used Marijuana}  & 22-25    & 0.72 & 81.1       \\
                               & 25-30     & 0.72 & 63.1       \\
                               & 30-37     & 0.71 & 65.9       \\ \bottomrule
\end{tabular}
\label{table:age_group}
\end{table}

\begin{figure}
\centering
\includegraphics[width=0.45\textwidth]{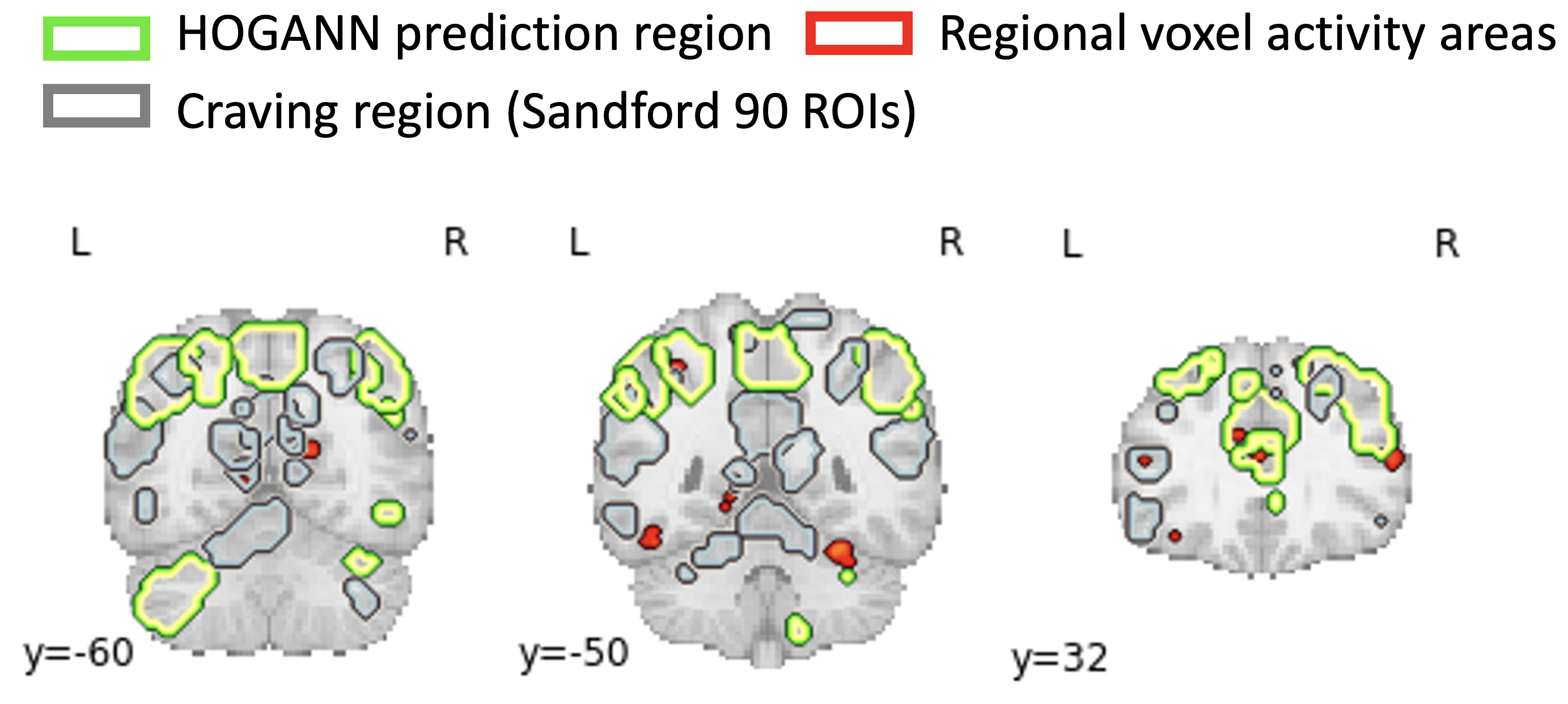}
\caption{Meta-analysis comparing HOGANN predictions with Neurosynth craving maps in marijuana-323. We superimposed 90 Stanford ROIs onto Neurosynth uniformity maps to identify craving-associated brain areas (shown in gray). The 3D voxel and average signal overlap illustrates regions of significant activation, with an applied z-score threshold (depicted in red) highlighting areas of substantial brain activity across studies focused on the keyword ‘craving’. The green highlighted areas indicate regions where HOGANN highly accurately predicted degree centrality, showing significant overlap with Stanford ROIs associated with chronic marijuana use and craving-related regions.}
\label{fig:predict_region.png}
\end{figure}

\begin{figure}
\centering
\includegraphics[width=0.8\textwidth]{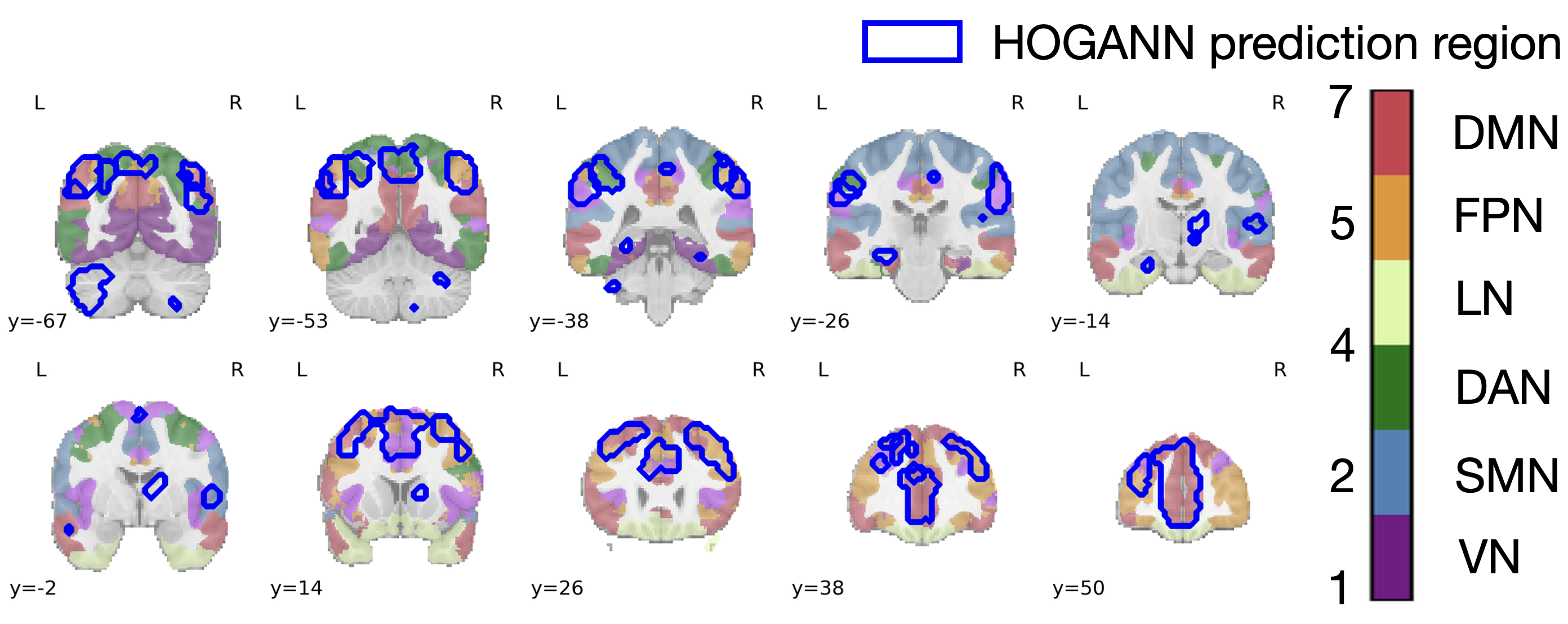}
\caption{The HOGANN predicts a seven-subnetwork parcellation of Yeo in the Marijuana-323.}
\label{fig:Yeo_7network}
\end{figure}

\begin{figure}
\centering
\includegraphics[width=0.8\textwidth]{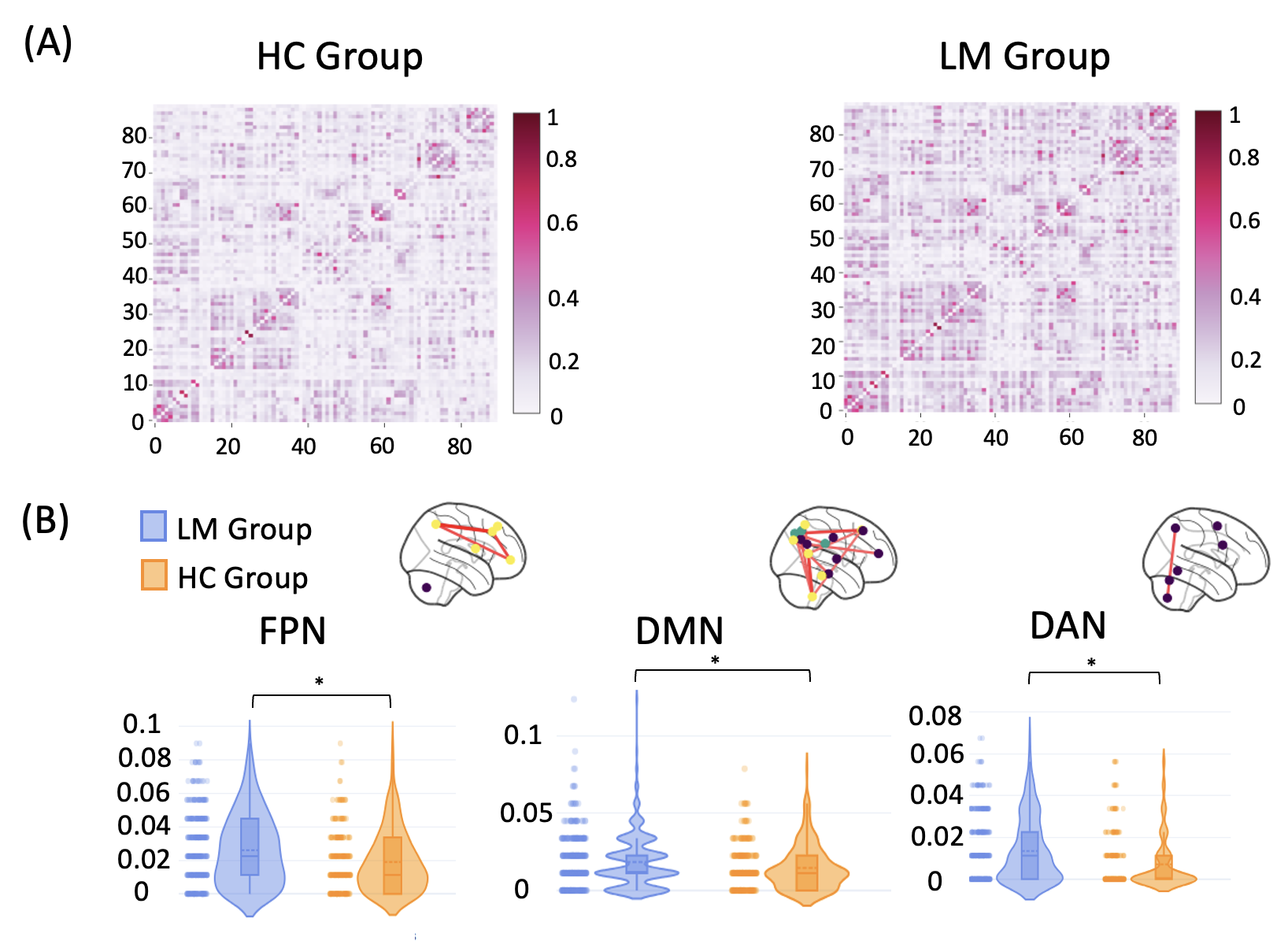}
\caption{The average of predicted functional weighted connectivity (WC) and degree centrality (DC) values ranking between ROIs in marijuana-323. Panel
(A) illustrates the comparison between the HC and LM groups, while  Panel (B) displays the corresponding network of the HC and LM groups, highlighting the regions with the highest DC values 
significant differences at paired sample $t$-test (``$\star$'' denotes $p<$  0.05) in Marijuana-323.}
\label{fig:weighted_dc_connectivity}
\end{figure}

\subsection{Community clustering}

This study explores the community structure within brain networks, emphasizing the significance of clustering for each region based on AWFC matrix and removing small connectivity edges with thresholds lower than 2\%. Our analysis is to identify the intricate connections in the LM group brain network and find community partitions. This examination is facilitated by integrating the functional connectivity of each instance with the trainable weighted matrix derived from our HOGANN prediction. Furthermore, we can rank the modularity~\cite{girvan2002community,fortunato2010community} and DC by utilizing the AWFC matrix to identify distinct clusters and community regions and compare with ML and our proposed model as demonstrated in Fig. \ref{fig:community_cluster}. This comparison, illustrated in Fig. \ref{fig:community_cluster} (A), demonstrates the derivation of the AWFC matrix using linear methods. The matrix is obtained by averaging the weight coefficients from four linear machine learning models (LR+L1, LR+L2, SVM+L1, and SVM+L2) and then multiplying these averages coefficient weight across all subjects in out-of-sample testing to find community clusters. In contrast, our proposed HOGANN-predicted AWFC matrix, as shown in Fig. \ref{fig:community_cluster} (B), highlights the superiority of the HOGANN in achieving enhanced modularity and has more significant community clustering, and we find out the significant region, particularly in community detection, as substantiated by the results.

As depicted in Fig. \ref{fig:community_cluster} (C), we encircled squares based on the color of each brain cluster region, representing their high-density degree of centrality. We highlighted critical brain regions associated with four communities by identifying them and sorting the importance scores according to the average degree centrality values. This allowed us to pinpoint essential brain regions linked to four distinct communities. In Marijuana-323 dataset, we have followed the indices corresponding to 90 ROIs from Stanford atlas definition~\cite{shirer2012decoding} and calculate DC scores that can be found within the four communities, as shown in Fig. \ref{fig:community_cluster} (D). It reveals the importance of the rankings of these values, offering insights into the spatial relationships in functional connectivity across these brain regions.  In our section \ref{sec:Yeo} on brain analysis, we will further analyze and interpret the regions of Yeo’s seven subnetworks corresponding to the four communities in Marijuana-323 (90 ROIs) and HCP (22 ROIs).

\begin{figure*}
\centering
\includegraphics[width=1\textwidth]{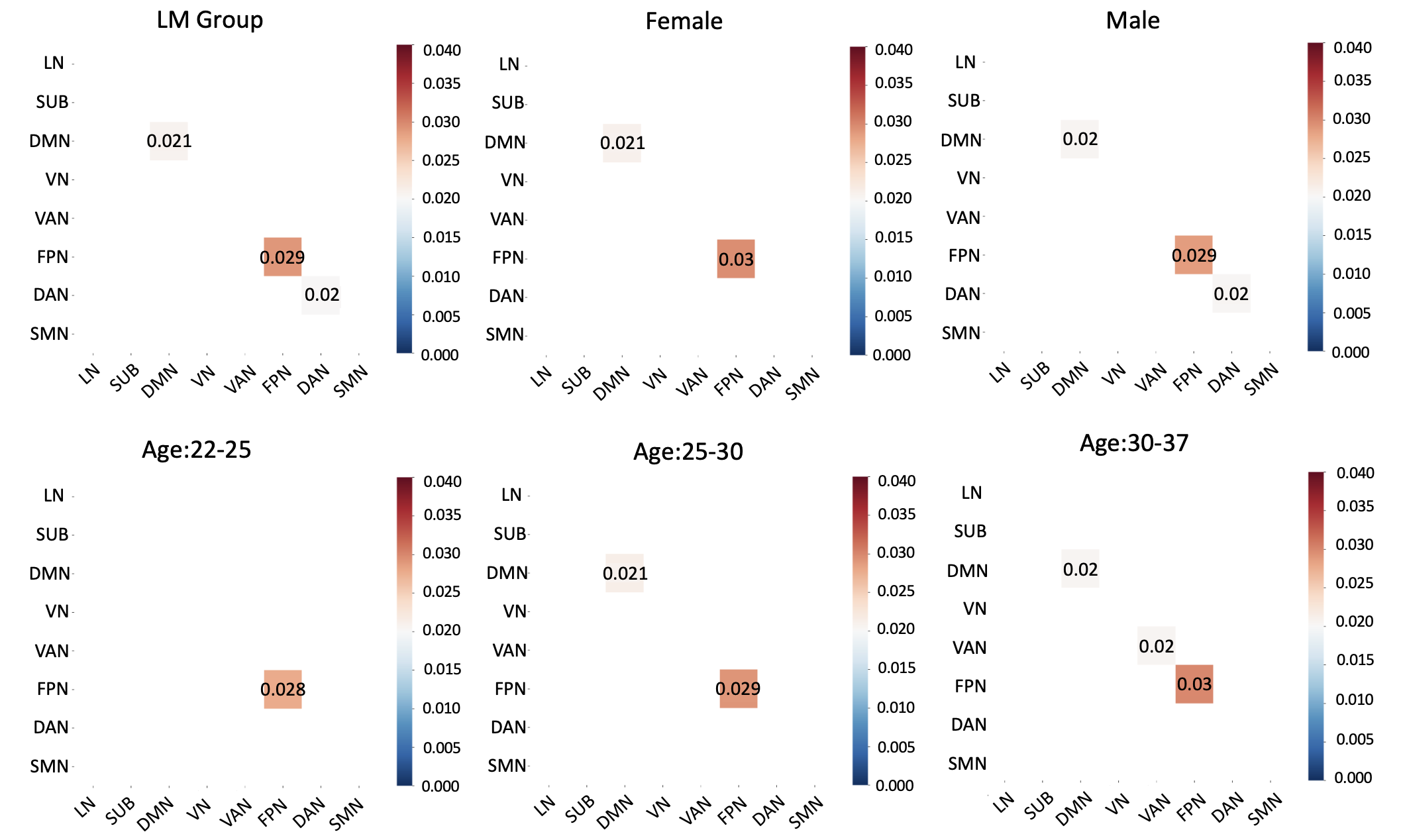}
\caption{Visualization of mean DC values across seven distinct networks in Marijuana-323, based on diversity analysis of LM groups.}
\label{fig:Age_network}
\end{figure*}

\begin{table}
\centering
\caption{Comparison of models predicting top-N Dice scores for keywords of Craving, Addiction, and Substance in target regions}
\setlength{\tabcolsep}{3pt} 
\begin{tabular}{l*{9}{>{\centering\arraybackslash}m{1.35cm}}}
\toprule
\textbf{Meta-Analysis} & \multicolumn{3}{c}{\shortstack{Craving \\ (num. studies=80)}} & \multicolumn{3}{c}{\shortstack{Addiction \\ (num. studies=135)}} & \multicolumn{3}{c}{\shortstack{Substance \\ (num. studies=124)}} \\
\cmidrule(lr){2-4} \cmidrule(lr){5-7} \cmidrule(lr){8-10}
& \textbf{Top-15 Dice} & \textbf{Top-20 Dice} & \textbf{Top-30 Dice} & \textbf{Top-15 Dice} & \textbf{Top-20 Dice} & \textbf{Top-30 Dice} & \textbf{Top-15 Dice} & \textbf{Top-20 Dice} & \textbf{Top-30 Dice} \\
\midrule
LR + L1 & \underline{0.27} & 0.29 & 0.36 & 0.18 & 0.26 & 0.33 & 0.31 & 0.27 & 0.34 \\
LR + L2 & 0.21 & 0.24 & 0.32 & 0.18 & 0.26 & 0.33 & 0.30 & 0.26 & \underline{0.38} \\
SVM + L1 & 0.21 & \underline{0.30} & 0.37 & 0.19 & 0.28 & 0.34 & 0.31 & 0.27 & 0.34 \\
SVM + L2 & 0.21 & 0.24 & 0.33 & 0.13 & 0.22 & 0.30 & 0.26 & 0.22 & 0.30 \\
XGboost & \underline{0.27} & 0.28 & \underline{0.40} & \underline{0.19} & \underline{0.32} & \underline{0.38} & \underline{0.32} & 0.28 & 0.30 \\
\hline
HOGANN & 0.39 & 0.44  & 0.48 & 0.22  & 0.35 & 0.40 & 0.33 & 0.34 & 0.47 \\
\bottomrule
\end{tabular}
\label{table:craving_addiction_substance}
\end{table}

\begin{table}
\centering
\caption{Brain regions categorized by affiliated Functional subnetwork.}
\begin{tabular}{|>{\centering\arraybackslash}m{1.5cm}|>{\centering\arraybackslash}m{3cm}|>{\centering\arraybackslash}m{3.5cm}|>{\centering\arraybackslash}m{4cm}|>{\centering\arraybackslash}m{4cm}|}
\hline ROIs && Frontoparietal Network (FPN)  &  Default Mode Network (DMN) &  Dorsal Attention Network (DAN) \\ \hline
\multirow{10}{*}{Stanford} 

 & Top-5 region & R middle frontal/dlPFC, L middle frontal/dlPFC & bilateral ACC & L inferior parietal/angular gyrus \\ \cline{2-5} 
 & Top-10 region & R middle frontal/dlPFC, L middle frontal/dlPFC & bilateral ACC, bilateral medial posterior precuneus, bilateral mPFC, bilateral posterior cingulate & L inferior parietal/angular gyrus, R inferior parietal/angular gyrus \\ \cline{2-5} 
 & Top-15 region & R middle frontal/dlPFC, L middle frontal/dlPFC, R middle frontal gyrus, L inferior frontal gyrus & bilateral ACC, bilateral medial posterior precuneus, bilateral mPFC, bilateral posterior cingulate & L inferior parietal/angular gyrus, R inferior parietal/angular gyrus \\ \hline
\multirow{11}{*}{Glasser} 
 
 & Top-5 region & Orbital\_and\_Polar\_Frontal, Inferior\_Frontal & Medial\_Temporal, Lateral\_Temporal, Posterior\_Cingulate & Primary\_Visual, Dorsal\_Stream\_Visual, MT  Complex and Neighboring Visual\_Areas \\ \cline{2-5} 
 & Top-10 region & Orbital\_and\_Polar\_Frontal, Inferior\_Frontal, Dorsolateral\_Prefrontal & Medial\_Temporal, Lateral\_Temporal, Posterior\_Cingulate,
 Anterior\_Cingulate,\
 Medial\_Prefrontal & \\ \cline{2-5} 
 & Top-15 region & Orbital\_and\_Polar\_Frontal, Inferior\_Frontal, Dorsolateral\_Prefrontal & Medial\_Temporal, Lateral\_Temporal, Posterior\_Cingulate, 
 
 Anterior\_Cingulate,\
 Medial\_Prefrontal & \\ \hline
\end{tabular}
\label{table:top_atlas}
\end{table}


\subsection{Craving maps characterization}

Craving maps can aid in deciphering brain network alterations in LM users. We utilized the Neurosynth package (\textbf{\url{https://neurosynth.org/}}), aligning it with the ”craving” keyword to identify consistently active brain regions across studies strongly associated with the term “craving”~\cite{yarkoni2011large}. Leveraging the design of the cited study, an automated meta-analysis of 80 publications identified regions activated by
 ”craving,” and Stanford ROIs can be mapped to the uniformity map~\cite{kulkarni2023interpretable}. Specifically, the uniformity map measures low-activity voxels by applying a threshold of 25\% during processing and identifies the activity in each brain region. The identified ROIs on the resulting map can be marked in red as threshold activity areas, as shown in Fig. \ref{fig:predict_region.png}. These ROIs represent areas substantially associated with craving, offering a comprehensive view of the brain’s response to LM use.

Next, we utilized the HOGANN to predict craving regions based on degree centrality (depicted by the green circle). We use uniformity maps to compare these predicted regions to the Neurosynth meta-analytic database. Finally, we invert the y-axis on the brain volumes within the Stanford ROIs, as indicated by the gray circle. The outcomes shown in  Fig. \ref{fig:predict_region.png} demonstrate the concordance of our HOGANN in predicting green cravings that are closely aligned with the ROIs across the axes at y = -60, y = -50 and y = 32. Our results showcase a successful analysis of cognitive-behavioral regions in LM users by effectively predicting specific activity regions. The prediction results through our model improve the prediction consistency and concordance of predictions between the network region and the craving maps from the meta-analysis of existing studies.

\subsection{Alterations in the intrinsic functional connectivity}\label{sec:Yeo}

Yeo et al.~\cite{yeo2011organization} identified seven intrinsic functional brain networks of the cerebral cortex using the cluster approach based on fMRI data from 1000 participants~\cite{yeo2011organization}. We utilized Yeo’s atlas of seven brain networks and analyzed the signal changes among these networks for SUD patients and normal controls. The seven networks include: 1. Visual Network (VN) 2. Sensorimotor Network (SMN) 3. Dorsal Attention Network (DAN) 4. Ventral Attention Network (VAN)
 5. Limbic Network (LN) 6. Frontoparietal Network (FPN) 7. Default Mode Network (DMN) 8. Subcortical System (SUB).
 In Fig. \ref{fig:Yeo_7network}, we plotted a uniformity map based on Yeo 7 resting-state subnetworks and marked regions predicted by HOGANN (in blue). In the first row, the green regions correspond to overlapping subnetworks at the top of the atlas, which are areas of the DAN associated working memory task \cite{brissenden2016functional}. In the second row, red regions represent DMN (attention and cognitive control). With Yeo’s seven subnetworks, we found the HOGANN could identify LM users with substantial regions in an active transition of cognitive function and attention.
 
According to our study’s observations, we align with previous relevant research that demonstrates LM users exhibit increased subcortical hyperconnectivity, leading to a widespread increase in functional connectivity across the whole brain connectome, particularly in the DMN and partially in the FPN and DAN~\cite{ramaekers2022functional,taebi2022shared}. Fig. \ref{fig:weighted_dc_connectivity} shows the 90 ROIs from the Stanford atlas definition ~\cite{shirer2012decoding} and the calculated DC matrices between the HC and LM group. This finding was corroborated by our analysis of the distinct DC matrices for the HC and LM groups in Fig. \ref{fig:weighted_dc_connectivity} (A), which revealed a higher density of functional activity regions in the LM group compared to the HC group. Notably, the LM group exhibited increased density in the FPN, DMN, and DAN networks, as illustrated in Fig. \ref{fig:weighted_dc_connectivity} (B). Moreover, this observation aligns with and is significantly supported by previous research on functional connectivity changes in networks among 2,390 addiction patients (1,143 with behavioral addiction and 1,456 with substance use disorder SUD) and 2,610 HC. These earlier studies consistently identified changes primarily involving the DMN and FPN~\cite{zeng2024similarity}.

\subsection{Craving region evaluation}

To better evaluate the overlap of craving region between HOGANN prediction and activate brain maps, the maps were thresholded at p$<$0.01 with FDR correction, and we can define the top-$N$ Dice score as: 

\begin{equation}
 Dice(N) = \frac{2 |Prediction(N) \cap Target(N)|}{|Prediction(N)| + |Target(N)|}
\end{equation}
where $|Prediction(N)|$ and $|Target(N)|$ represent the number of top-N ROIs in the predicted activation map and the ground truth ROIs (e.g., Stanford 90 ROIs). The $|Prediction(N)| + |Target(N)|$ present the overlap between predicted Top-N region and target ROIs. Observing the results in Table \ref{table:craving_addiction_substance}, we analyzed keywords such as Craving, Addiction, and Substance under HOGANN. We compared their Top-N-Dice scores with other baseline methods from the meta-analysis. HOGANN showed the second-highest Dice score. For the keyword ‘Craving’ HOGANN's Top-20 Dice score improved by 46.6\%. For the keyword "Addiction," the Top-15 Dice score improved by 15.7\%, and for "Substance," the Top-30 Dice score improved by 23.6\%. The results indicate that HOGANN can identify better overlapping regions in the craving regions associated with LM use.

\section{Discussion}

This study is the first to investigate the spatial craving maps among the LM users and the prediction of brain functional connectivity alterations based on two fMRI datasets. Our findings reveal several critical  insights into the brain's functional organization associated with marijuana use: (1) HOGANN predicted the results more accurately for individuals with marijuana dependence versus marijuana users, demonstrating that these changes in brain function were linked to severity of use, (2) we used HOGANN to predict mean DC levels in different atlas regions (e.g., 90 and 22 ROIs) across the two datasets, focusing on the most frequently associated regions within Yeo's seven networks. Notably, among the Top-N regions in these two atlases, the most frequently shared affiliated functional subnetworks were the FPN, DMN, and DAN, as shown in Table \ref{table:top_atlas}, (3) by comparing Yeo's seven subnetworks to the four communities in Fig. \ref{fig:community_cluster} (D), community 1, 3 and 4 mainly involve the DMN, while community 2 and mainly involve the FPN. These results align with the changes in functional connectivity caused by nicotine use~\cite{wetherill2015cannabis} and the significant impact on the DMN regions caused by marijuana administration~\cite{whitfield2018understanding}, (4) compared to other ML methods, HOGANN demonstrated the highest Dice score for the keyword 'Craving' among all keywords analyzed.

Several neurological studies have examined the treatment of marijuana addiction and its effects on three crucial brain networks including FPN, DMN, and DAN. These networks consistently show enhanced activity during drug-related cue processing but diminished activity in non-drug-related cognitive tasks \cite{zilverstand2018neuroimaging}. FPN and DAN abnormalities appear to span across various stages of addiction, with FPN activity positively correlating with addiction severity, craving levels, and relapse risk, while DAN activity is associated with frequency of drug use. DMN abnormalities, on the other hand, are primarily associated with subjective craving. Importantly, the FPN is a crucial hub among brain networks~\cite{rawls2022resting} and therapeutic interventions have shown potential in upregulating FPN activity for addiction treatment~\cite{song2022reducing,song2019effects}.

Moreover, we further examined how different age cohorts changed across time periods. Our analysis revealed that individuals aged 25-30 displayed greater mean DC in the DMN compared to the 22-25 age group. Notably, older participants demonstrated more pronounced alterations in both the DMN and VAN over time, as  illustrated in Fig. \ref{fig:Age_network}. These findings suggest a positive relationship between advancing age and functional connectivity \cite{harding2012functional,watson2022cannabis,lichenstein2022systematic,wang2019abnormal}.

\section{Conclusion}

In this study, we propose a mixed model (HOGANN) that incorporates a high-order attention mechanism to enhance the neighboring connection features information of fMRI time series within topological graph structures. The HOGANN demonstrates superior accuracy in classifying multigraphs of brain networks and identifying overlapping craving regions. Furthermore, by utilizing the HOGANN in conjunction with the Yeo atlas of seven intrinsic networks, we successfully validated common significant subnetworks across two datasets in LM groups. By identifying specific network alterations associated with marijuana use, our study contributes to the growing body of evidence supporting a multi-faceted, network-based understanding of addiction.  Additionally, we effectively elucidated how different demographic factors in the LM group correlate with varying connectivity patterns in brain network craving regions. These findings provide valuable insights that lead to more targeted interventions aimed at normalizing these network dysfunctions in research and treatment of marijuana withdrawal.

\bibliographystyle{unsrt}  
\bibliography{references}

\end{document}